\documentclass[10pt,preprint]{aastex}

\usepackage{graphicx,apjfonts,emulateapj5,onecolfloat5}
\usepackage{subfigure}
\usepackage{ulem}

\shorttitle{NATURE OF W51e2: MASSIVE CORES AT DIFFERENT PHASES OF STAR
  FORMATION} 
\shortauthors{SHI, ZHAO, \& HAN}

\begin{document}
\twocolumn[


\title{NATURE OF W51e2: MASSIVE CORES AT DIFFERENT PHASES OF STAR
  FORMATION}


\author{Hui Shi\altaffilmark{1},
        Jun-Hui Zhao\altaffilmark{2},
        J.L. Han\altaffilmark{1},
        }


\begin{abstract}
We present high-resolution continuum images of the W51e2 complex
processed from archival data of the Submillimeter Array (SMA) at 0.85
and 1.3\,mm and the Very Large Array at 7 and 13\,mm. We also made
line images and profiles of W51e2 for three hydrogen radio
recombination lines (RRLs; H26$\alpha$, H53$\alpha$, and H66$\alpha$)
and absorption of two molecular lines of HCN(4-3) and CO(2-1). At
least four distinct continuum components have been detected in the
3$\arcsec$ region of W51e2 from the SMA continuum images at 0.85 and
1.3\,mm with resolutions of 0.3$\arcsec\times0.2\arcsec$ and
1.4$\arcsec\times0.7\arcsec$, respectively.
The west component, W51e2-W, coincides with the ultracompact HII
region reported from previous radio observations. The H26$\alpha$ line
observation reveals an unresolved hyper-compact ionized core
($<0.06\arcsec$ or $<310$\,AU) with a high electron temperature of
$1.2\times10^4$\,K, with the corresponding emission measure EM$>7
\times10^{10} {\rm pc~cm^{-6}}$ and the electron density $N_e>7
\times10^6$ cm$^{-3}$. The inferred Lyman continuum flux implies that
the HII region W51e2-W requires a newly formed massive star, an O8
star or a cluster of B-type stars, to maintain the ionization.
W51e2-E, the brightest component at 0.85\,mm, is located 0.9$\arcsec$
east from the hyper-compact ionized core. It has a total mass of
$\sim$140\,M$_{\sun}$ according to our spectral energy distribution
analysis and a large infall rate of $>
1.3\times10^{-3}$\,M$_{\sun}$yr$^{-1}$ inferred from the absorption of
HCN.  W51e2-E appears to be the accretion center in W51e2. Given the
fact that no free$-$free emission and no RRLs have been detected, the
massive core of W51e2-E appears to host one or more growing massive
proto-stars.
Located $2\arcsec$ northwest from W51e2-E, W51e2-NW is detected in the
continuum emission at 0.85 and 1.3\,mm. No continuum emission has been
detected at $\lambda\ge$ 7\,mm. Along with the maser activities
previously observed, our analysis suggests that W51e2-NW is at an
earlier phase of star formation.
W51e2-N is located 2$\arcsec$ north of W51e2-E and has only been
detected at 1.3\,mm with a lower angular resolution ($\sim1\arcsec$),
suggesting that it is a primordial, massive gas clump in the W51e2
complex.

\end{abstract}


\keywords{HII regions -- ISM: individual objects (W51e2) -- stars:
  formation} ]

\altaffiltext{1}{National Astronomical Observatories, Chinese Academy
  of Sciences, Jia20 Datun Road, Chaoyang District, Beijing 100012,
  China.  Email: shihui, hjl @ nao.cas.cn}

\altaffiltext{2}{Harvard-Smithsonian Center for Astrophysics, 60
  Garden Street, Cambridge, MA 02138, USA. Email: jzhao @
  cfa.harvard.edu}


\section{INTRODUCTION}
Massive stars are formed in dense, massive molecular cores.  Detailed
physical processes in star-forming regions have not been well studied
until recent high-resolution observations were available at
submillimeter wavelengths. High angular resolution observations are
necessary to unveil the physical environs and activities of the
individual sub-cores as well as their impact on the overall process of
massive star formation.

W51 is a well-known complex of HII regions with about $1\degr$ area in
the Galactic plane \citep{Wes58}. W51A, also called G49.5-0.4, is the
most luminous region in W51 \citep{KV67}. Several discrete components
have been found from W51A and referred alphabetically as a$-$i in the
right ascension (R.A.) order \citep{Mar72,Meh94}. W51e is one of the
brightest regions in W51A.  High-resolution centimeter observations
have revealed that W51e consists of several ultracompact (UC) HII
regions \citep{Sco78,GJW93}, among which the UC HII region of W51e2
appears to be the brightest.  This region is severely obscured by the
dust, and no IR detection has been made at wavelengths shorter than
20\,$\mu$m \citep{GBM82}. The inferred distance to W51e2 is
$\sim$5.1\,kpc \citep{XRM09}; thus, 1$\arcsec$ corresponds to a linear
scale of about 5100\,AU.

Based on the radio continuum observations, spectral energy
distribution (SED) analyses of W51e2 suggest that the emission source
from this region consists of two components, namely, (1) thermal
emission from a cold dust component and (2) free$-$free emission from
a hot HII region \citep{RWP90, SZH04, KZK08}.

Observations of hydrogen radio recombination lines (RRLs) were carried
out toward W51e2 \citep{Meh94,KZK08,KK08}. \citet{Meh94} failed to
detect the H92$\alpha$ line in this region, which might be attributed
to the large optical depth and/or the large pressure broadening of the
UC HII region at 3.6\,cm. Based on the observations of H66$\alpha$,
H53$\alpha$, and H30$\alpha$, \citet{KZK08} argued that the pressure
broadening is responsible for the broad-line widths observed in the
RRLs.

\begin{table*}
\small
\centering
\tablewidth{3pt}
\setlength{\tabcolsep}{0.8mm}
%
\caption{Observation parameters for W51e2}
\label{tab:observation}
%
\begin{tabular}{lllll}
\hline
\hline
\multicolumn{1}{l}{Parameters} &
\multicolumn{1}{l}{SMA: 0.85\,mm} &
\multicolumn{1}{l}{SMA: 1.3\,mm} &
\multicolumn{1}{l}{VLA: 7\,mm} &
\multicolumn{1}{l}{VLA: 13\,mm} \\
%
%
\hline
Observation date     &2007 Jun. 18  &2005 Sep. 1  &2004 Feb. 14 &2003 Sep. 29\\
Array configuration  &Very extended (8 ants)&Extended (6 ants) &AB   &BC \\
Pointing center R.A.(J2000)&19:23:43.888  &19:23:43.895  & 19:23:43.918 &19:23:43.918 \\
Pointing center decl.(J2000)&+14:30:34.798&+14:30:34.798 &+14:30:28.164&+14:30:28.164\\
Frequency(GHz)   &343, 353&221, 231&43             &22            \\
Bandwidth        &2.0+2.0 (GHz)      &2.0+2.0 (GHz)      &12.5 (MHz)    &12.5 (MHz) \\
On-source time (hr) &2.10           &4.75          &5.90         &3.93        \\
System temperature(K)&150$\sim$500  &100$\sim$250  &             &            \\
Bandpass calibrators &J1229+020, J1751+096& 3C 454.4&3C 84        &3C 84             \\
Phase calibrators    &J1733$-$130, J1743$-$038&J1751+096, J2025+337&J1923+210     &J1923+210\\
                     &J2015+371              &                   &              &\\
Flux calibrators     &Callisto      &Ceres         &3C 286       &3C 286            \\
\hline
\end{tabular}
\end{table*}

Observations of molecular lines \citep[e.g.][]{HY96, ZH97, ZHO98,
  YKH98, SZH04} showed evidence for infall (or accretion) gas within
5$\arcsec$ ($<$ 0.2\,pc) around W51e2. A possible rotation was
suggested by \citet{ZH97} on the basis of fitting the
position$-$velocity (PV) diagram of the NH$_3$(3,3) absorption line at
13\,mm.  A spin-up rotation with an axis of position angle (P.A.)
$\approx$20$\degr$ was further suggested based on the velocity
gradient of CH$_3$CN(8$_3$$-$7$_3$) at 2\,mm \citep{ZHO98}. However,
based on the velocity gradient observed from the H53$\alpha$ line,
\citet{KK08} interpreted their results as evidence for rotational
ionized accretion flow around the UC HII region and derived a
rotational axis of P.A.$\approx-30\degr$. In addition, a possible
bipolar outflow along the northwest and southeast (NW$-$SE) directions
was suggested based on the observations of the CO(2-1) line emission
\citep{KK08}, which appears to be perpendicular to the axis of the
rotational ionized disk. However, the southwest (SW) elongation of the
UC HII region in Gaume et al.'s (1993) observations was explained as a
collimated ionized outflow. On the other hand, proper-motion
measurements of H$_2$O/OH masers appeared to favor a hypothetical
model with an outflow or expanding gas bubble in W51e2
\citep{IWO02,FR07}.

Thus, improved images and comprehensive analysis of high-resolution
observations at the wavelengths from radio to submillimeter are needed
to reconcile the differences in the interpretation of the results from
the previous observations of W51e2. In this paper, we show the
high-resolution continuum images with the Submillimeter Array (SMA) at
0.85 and 1.3\,mm and with the Very Large Array (VLA) at 7 and 13\,mm,
as well as line images and profiles of the hydrogen recombination
lines (H26$\alpha$, H53$\alpha$, and H66$\alpha$) and the molecular
lines of HCN(4-3) and CO(2-1). In Section 2, we describe the details
of the data processing and results.  In Section 3, we model the
individual components on the basis of the high-resolution observations
of W51e2 with the SMA and VLA.  The nature and astrophysical processes
of star formation cores in W51e2 are discussed. A summary and
conclusions are given in Section 4.

\section{DATA REDUCTION AND RESULTS}
We processed the archival data from the SMA and VLA observations of
W51e2 and constructed both the continuum and spectral line
images. Details in calibration and imaging are discussed below along
with the relevant parameters that are summarized in Table
\ref{tab:observation}.

\subsection{The SMA continuum images at 0.85 and 1.3\,mm}
The observations of W51e2 at 0.85\,mm were carried out with the SMA in
the very extended array (eight antennas) and at 1.3\,mm in the
extended array (six antennas). The data were calibrated in Miriad
\citep{STW95} following the reduction instructions for SMA data
\footnote{\it http://www.cfa.harvard.edu/sma/miriad}. The system
temperature corrections were applied to the visibility data.  The
bandpass calibration was made with strong calibrators for each of the
two data sets (Table \ref{tab:observation}). The flux density scale
was determined using Callisto at 0.85\,mm and Ceres at 1.3\,mm.

Corrections to the complex gains at 1.3\,mm were made by applying the
solutions interpolated from the two nearby QSOs J1751+096 and
J2025+337. For the high-resolution data at 0.85\,mm, the visibility
data are initially calibrated with a model of W51e2 derived from
observations in a low angular resolution at 0.87\,mm. The residual
phase errors were further corrected using the self-calibration
technique.  The final continuum images were constructed by combining
all the line-free channels in both upper sideband and lower sideband
data.

\begin{figure}[t]
\centering
  \includegraphics[angle=270,width=90mm]{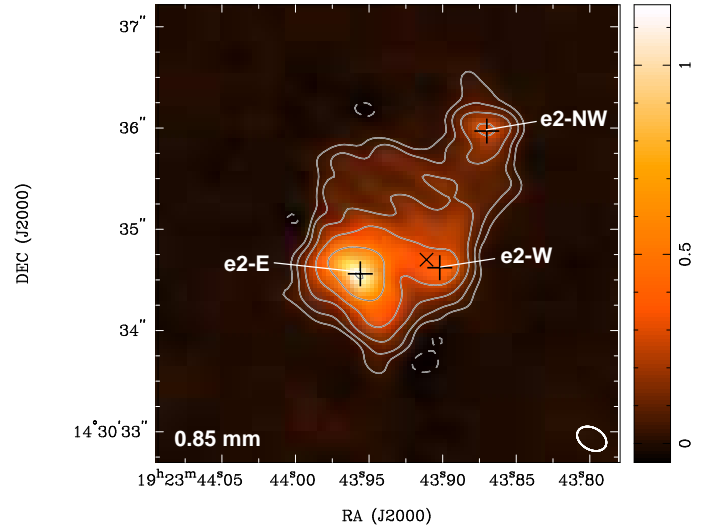}
\caption{
High-resolution 0.85 mm continuum image of W51e2 observed with the
SMA. Contours are $\pm 5\sigma \times 2^n$ ($n=0$, 1, 2, 3, ... and
$\sigma=7.11$~mJy~beam$^{-1}$).  The FWHM beam ($0.31\arcsec \times
0.22\arcsec$, P.A.=$60.6\degr$) is shown at bottom right.  Three
continuum emission components e2-NW, e2-E, and e2-W are marked with
``+''.  The $\lambda3.6$\,cm position of the UC HII region in W51e2
\citet{GJW93} is marked with ``$\times$''.
}
\label{fig:350GHz}
\end{figure}

\begin{figure*}[t]
\centering
  \subfigure[]{\includegraphics[angle=270,width=58mm]{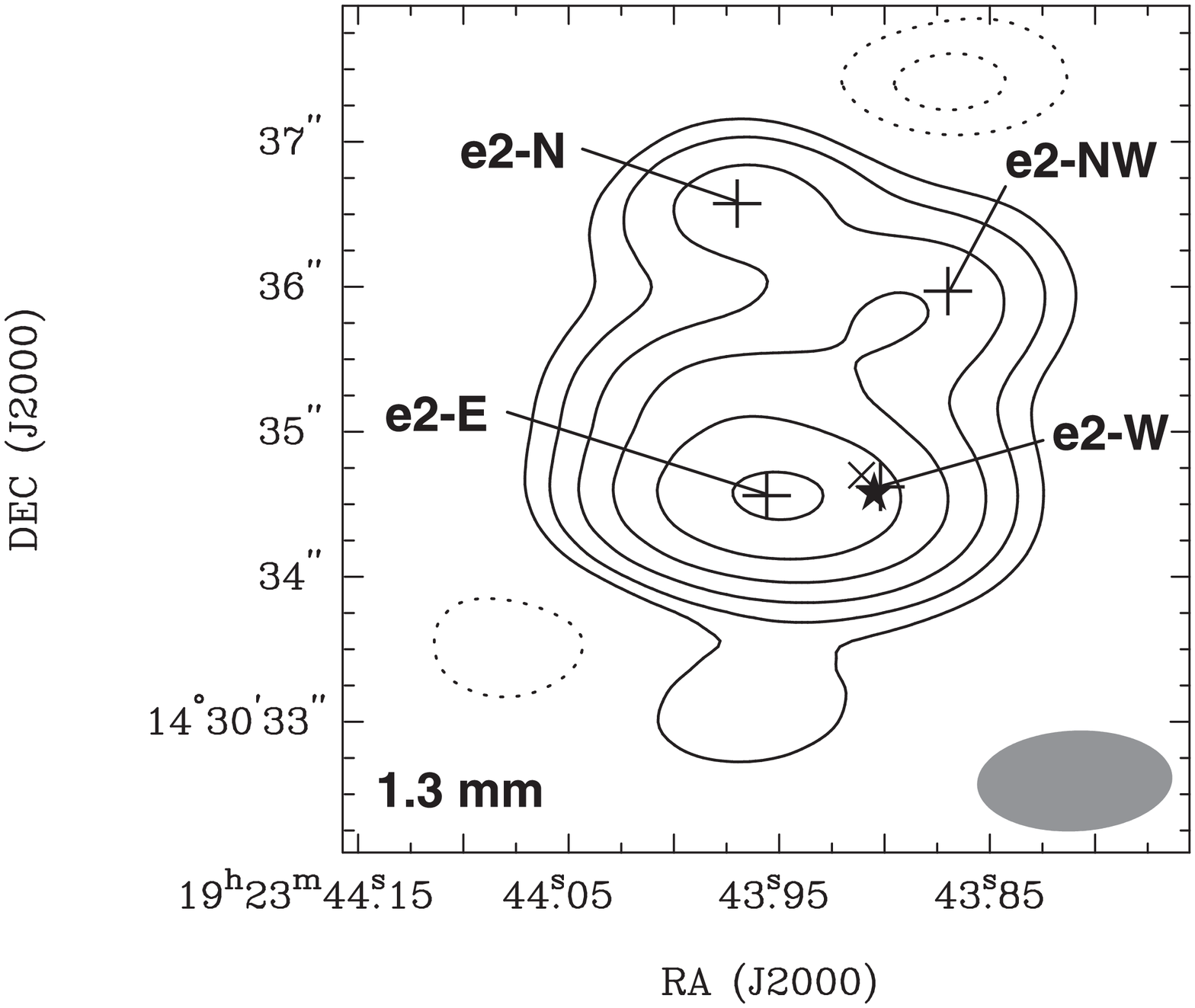}}
  \subfigure[]{\includegraphics[angle=270,width=58mm]{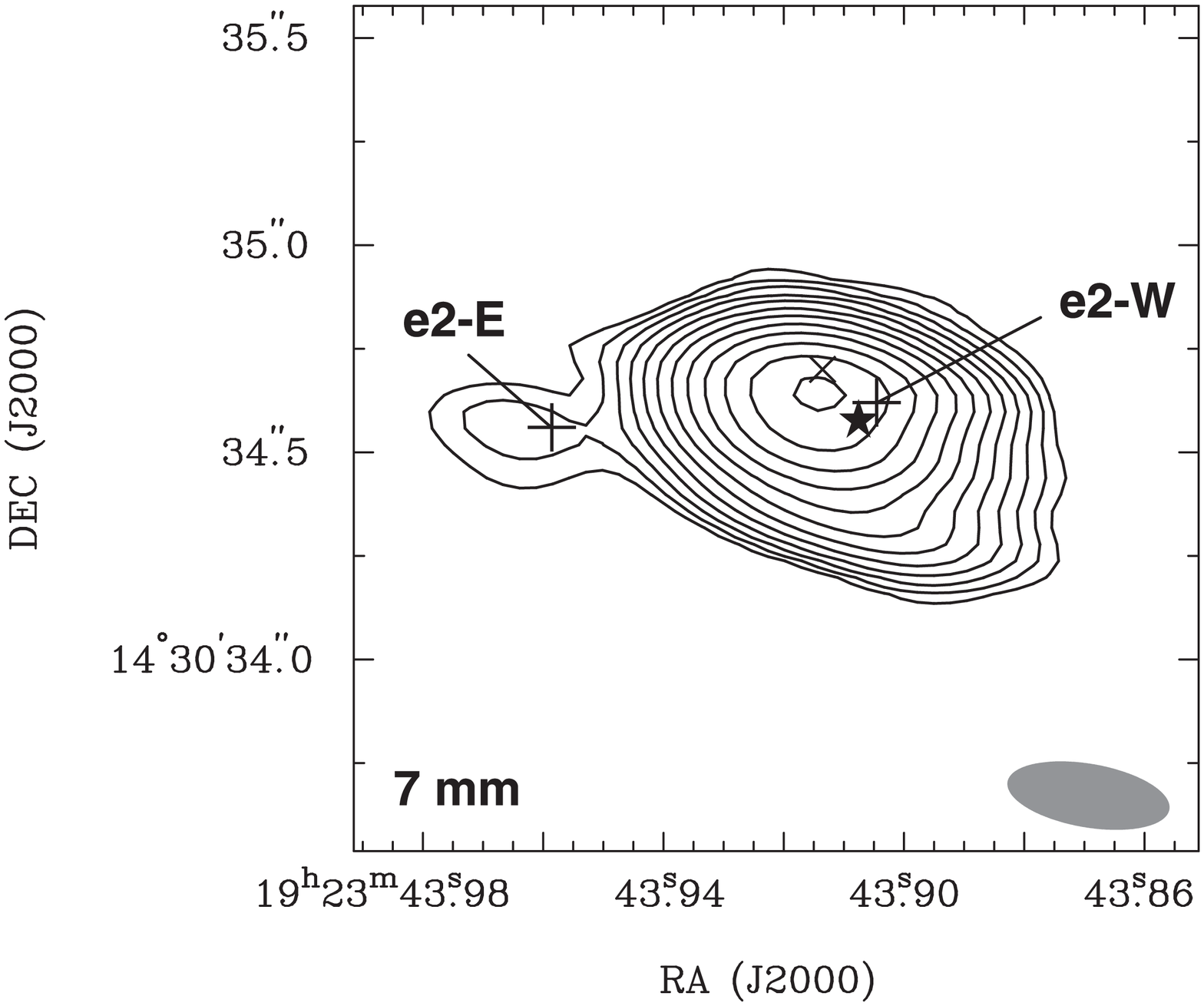}}
  \subfigure[]{\includegraphics[angle=270,width=58mm]{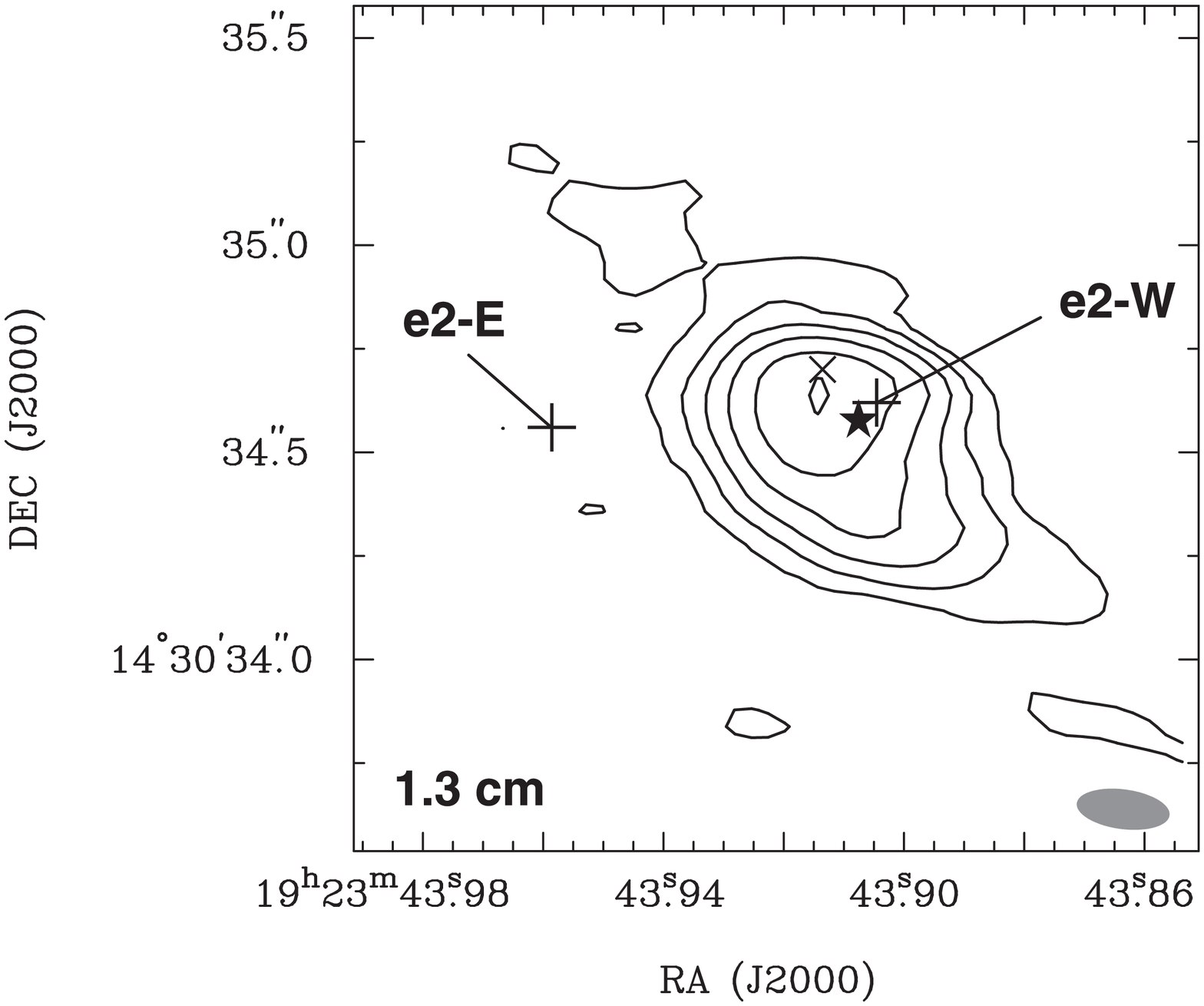}}\\
  \subfigure[]{\includegraphics[angle=270,width=58mm]{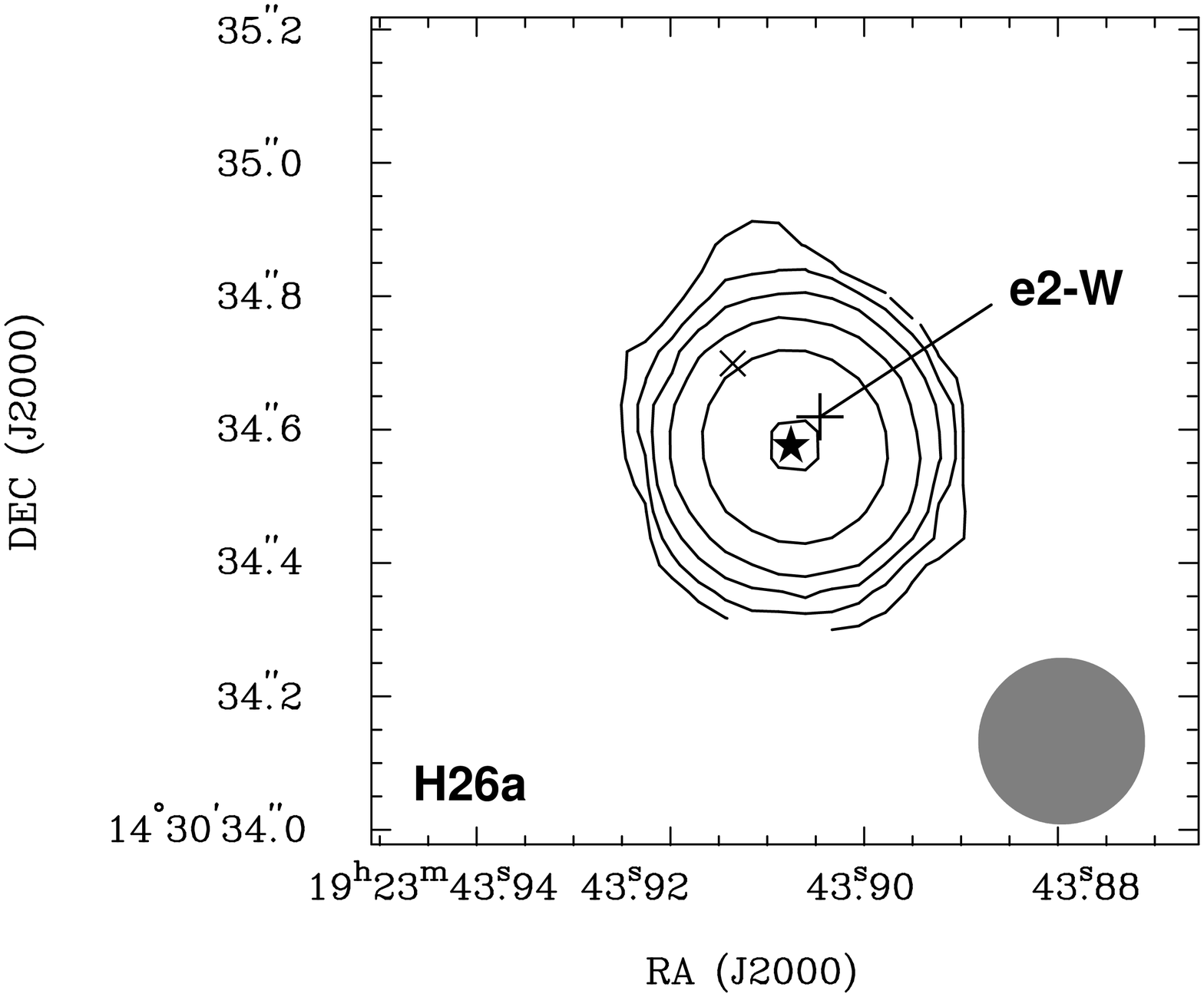}}
  \subfigure[]{\includegraphics[angle=270,width=58mm]{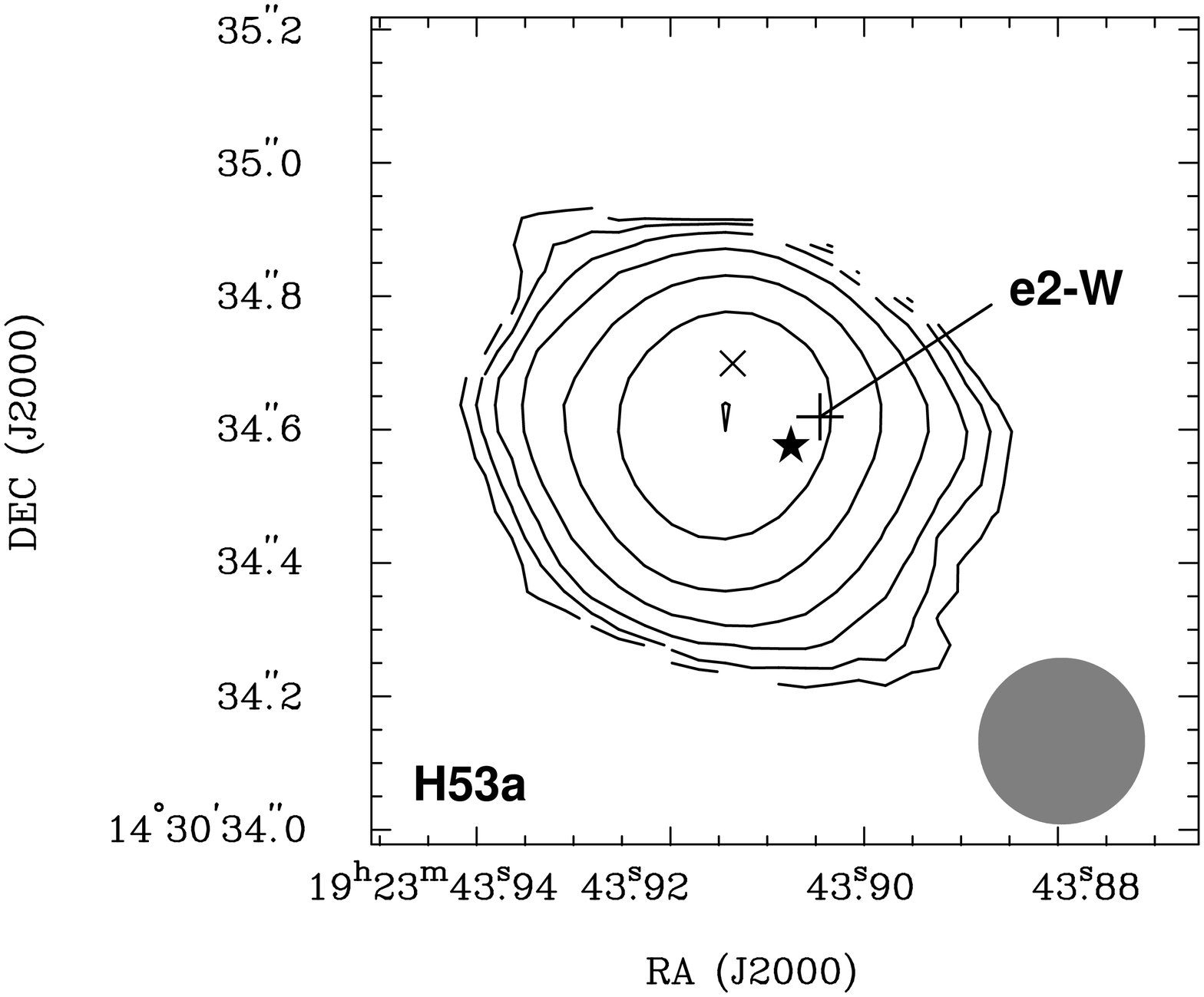}}
  \subfigure[]{\includegraphics[angle=270,width=58mm]{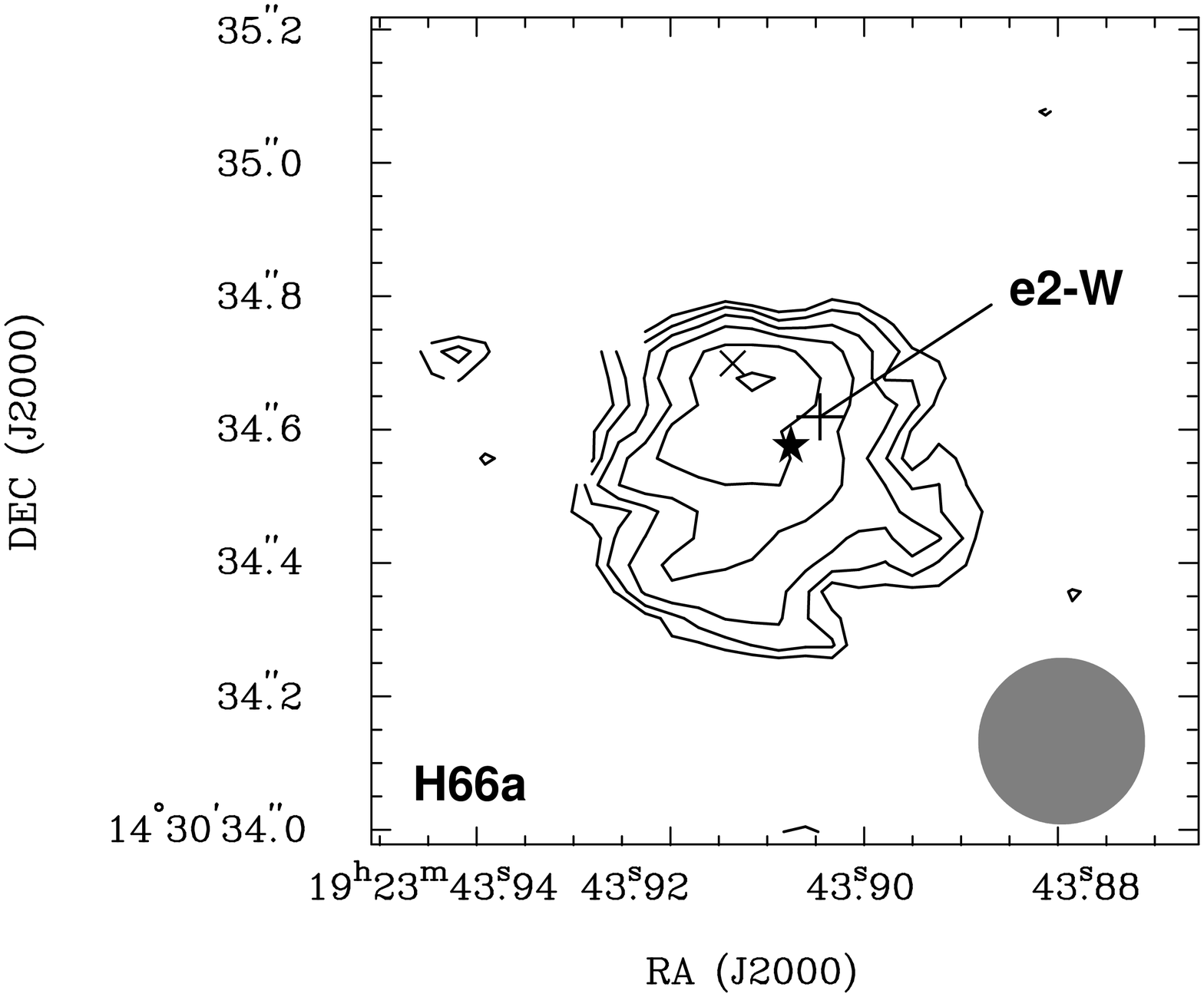}}
\caption{
{\it (a)}: 1.3\,mm continuum image of W51e2 observed with the
SMA. Contours are $\pm 5\sigma \times 2^n$ ($n=0$, 1, 2, 3, ... and
$\sigma=14.0$~mJy~beam$^{-1}$). The FWHM beam ($1.35\arcsec \times
0.69\arcsec$, P.A.=$-87.4\degr$) is shown at bottom right.  The three
components detected at 0.85\,mm together with the additional component
e2-N detected at 1.3\,mm are marked with ``+''.  The position ``+'' of
the submillimeter components and the peaks of the 3.6\,cm continuum
(``$\times$'') and H26$\alpha$ (``$\bigstar$'') are also marked in the
rest of the images.
{\it (b)}: 7\,mm continuum image of W51e2 observed with the
VLA. Contours are $\pm 5\sigma \times 2^{n/2}$ ($n=0$, 1, 2, 3,
... and $\sigma=1.12$~mJy~beam$^{-1}$).  The FWHM beam ($0.40\arcsec
\times 0.15\arcsec$, P.A.=$80.3\degr$) is shown at bottom-right.
{\it (c)}: 13\,mm continuum image of W51e2 observed with VLA.
Contours are $\pm 5\sigma \times 2^n$ ($n=0$, 1, 2, 3, ..., and
$\sigma=0.50$~mJy~beam$^{-1}$). The FWHM beam ($0.23\arcsec \times
0.10\arcsec$, P.A.=$82.6\degr$) is shown at bottom right.
{\it(d)}: Image of the H26$\alpha$ line emission integrated from
25\,km~s$^{-1}$ to 84\,km~s$^{-1}$ from W51e2.  Contours are $\pm
5\sigma \times 2^n$ ($n=0$, 1, 2, 3, ... and
$\sigma=0.146$~Jy~beam$^{-1}$~km~s$^{-1}$).  The FWHM beam
($0.25\arcsec$) is shown at bottom right. All the RRL images are
convolved to a common circular beam ($0.25\arcsec$).  The star denotes
the peak position of the H26$\alpha$ line and is marked in the rest of
the RRL images.
{\it(e)}: Image of the H53$\alpha$ line emission integrated from
22\,km~s$^{-1}$ to 101\,km~s$^{-1}$. Contours are $\pm 5\sigma \times
2^n$ ($n=0$, 1, 2, 3, ... and
$\sigma=9.68\times10^{-3}$~Jy~beam$^{-1}$~km~s$^{-1}$).
{\it(f)}: Image of the H66$\alpha$ line emission integrated from
24\,km~s$^{-1}$ to 120\,km~s$^{-1}$. Contours are $\pm 5\sigma \times
2^{n/2}$ ($n=0$, 1, 2, 3, ... and
$\sigma=6.14\times10^{-3}$~Jy~beam$^{-1}$~km~s$^{-1}$).
}
\label{fig:con_rrls}
\end{figure*}

Figure \ref{fig:350GHz} shows the high-resolution
(0.3$\arcsec\times0.2\arcsec$) continuum image of W51e2 at
0.85\,mm. The complex of W51e2 has been resolved into at least three
bright, compact components (see Figure \ref{fig:350GHz}, the symbol
``+'' marks the positions of the components).  The brightest source,
W51e2-E, is located $\sim0.9\arcsec$ east of the UC HII region which
is marked with ``$\times$'' for the peak position at 3.6\,cm from
\citet{GJW93}. The secondary bright source, W51e2-W, coincides with
the 3.6\,cm peak of the UC HII region. The third component, W51e2-NW,
is located about 2$\arcsec$ northwest of W51e2-E. In addition, these
compact emission components appear to be surrounded by an amorphous
halo in the continuum emission at 0.85\,mm. The three bright continuum
sources were also detected in the low-resolution continuum image at
1.3\,mm (Figure \ref{fig:con_rrls}(a)). Moreover, a weak continuum
component, W51e2-N, is present $\sim2\arcsec$ north of W51e2-E.

\subsection{The VLA continuum images at 7 and 13\,mm}
The observations of W51e2 at 7 and 13\,mm were carried out with the
VLA (see the summarized observing parameters in Table
\ref{tab:observation}). We made the calibrations by following the
standard data-reduction procedure for VLA data with
AIPS\footnote{http://www.aips.nrao.edu}. Then, the calibrated
visibilities were loaded into the Miriad environment. The imaging and
further analysis were carried out with Miriad.

The dirty images were made using INVERT with robust weighting
(robust=0) and were cleaned with the hybrid deconvolution
algorithm. Figures \ref{fig:con_rrls}(b) and \ref{fig:con_rrls}(c)
show the continuum images at 7 and 13\,mm, respectively. Strong
continuum emission is detected from W51e2-W at both 7 and 13\,mm in
good agreement with previous observations at centimeters
\citep{GJW93}. In addition, a weak but significant ($>$10$\sigma$)
emission feature at W51e2-E has been detected at only 7\,mm. As shown
in Figure \ref{fig:con_rrls}(b), the weak 7\,mm continuum source at
W51e2-E has been clearly separated from the UC HII region, W51e2-W.

\subsection{Hydrogen recombination lines}
For the line cubes, the continuum emission was subtracted using the
linear interpolation from the line-free channels with UVLIN.  The
H26$\alpha$ line (353.623\,GHz) was included in the line cubes made
from the SMA data at 0.85\,mm. We also made a line image cube to cover
the H30$\alpha$ line (231.901\,GHz) from the SMA data at 1.3\,mm. Both
the H26$\alpha$ and H30$\alpha$ line data were resampled to
1\,km~s$^{-1}$. The rms noises of 49\,mJy beam$^{-1}$ and 81\,mJy
beam$^{-1}$ in each of the channel images are inferred for the
H26$\alpha$ and H30$\alpha$ line image cubes, respectively. Figure
\ref{fig:con_rrls}(d) shows the integrated line intensity image of the
H26$\alpha$ line made with a 3$\sigma$ cutoff in the velocity range
from 25 to 84\,km~s$^{-1}$. The H26$\alpha$ line emission from the UC
HII region (W51e2-W) appears to be very compact and has not been
resolved with the beam of 0.25$\arcsec$, suggesting that the intrinsic
source size of the hyper-compact HII core is $<0.06\arcsec$. The peak
position of the integrated H26$\alpha$ line emission (stars in Figure
\ref{fig:con_rrls}) appears to have a significant offset of
$\sim$0.15$\arcsec$ from the continuum peak at 13\,mm (also see the
Gaume's position in Figure \ref{fig:con_rrls}(d)). Unlike the other
RRLs and continuum emission at centimeter wavelengths, the line
emission at the frequency of the H30$\alpha$ line (the image is not
shown in this paper) shows an extended distribution covering the
entire 3$\arcsec$ region, suggesting that the H30$\alpha$ line might
have been severely contaminated by one or more molecular lines in
W51e2 (e.g., C$_3$H$_7$CN with a rest frequency of 231.9009\,GHz).

\begin{figure}[t]
\centering
\includegraphics[angle=270,width=80mm]{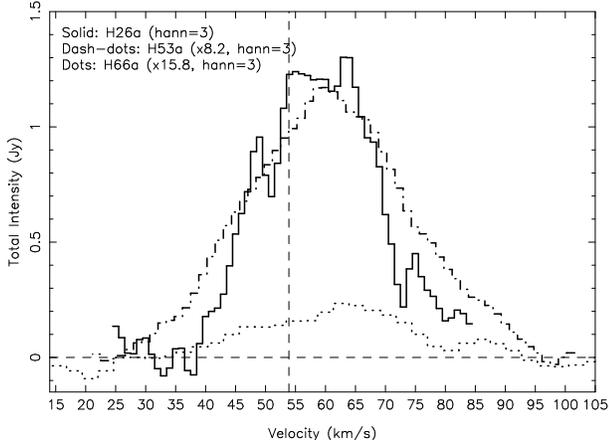}
\caption{
Profiles of hydrogen recombination lines, H26$\alpha$, H53$\alpha$,
and H66$\alpha$, integrated from W51e2-W in an area of
0.5$\arcsec\times$0.5$\arcsec$ centered at the H26$\alpha$ line
peak. All the line images have been convolved with the same circular
beam of 0.25$\arcsec$.  The spectra are multiplied with the ratios of
the rest frequency of the H26$\alpha$ line to the RRL rest frequencies
($\nu_{H26\alpha}/\nu_{RRL}$).  The vertical dashed line denotes the
systematic velocity of 53.9$\pm$1.1\,km~s$^{-1}$ which we determined
from the previous measurements of seven different hot molecular lines
(see Section 2.4).}
\label{fig:profiles}
\end{figure}

The line of H66$\alpha$ is broad and weak, which is barely covered by
the VLA correlator band with which we had difficulty determining the
line-free channels. Instead, the continuum levels were determined by
averaging all channels excluding a few relatively strong line emission
channels and were subtracted from the visibility data.  The
uncertainty is $<5$\% in the continuum levels due to the line
contamination. The image cubes of the H53$\alpha$ and H66$\alpha$
lines were cleaned and convolved to a circular beam of 0.25$\arcsec$,
giving the rms noises of 2.4 and 1.5\,mJy~beam$^{-1}$ per channel,
respectively.  The integrated intensities of H53$\alpha$ (with a
3$\sigma$ cutoff in each channel) and H66$\alpha$ (with a 2$\sigma$
cutoff in each channel) are shown in Figures \ref{fig:con_rrls}(e) and
\ref{fig:con_rrls}(f), respectively. Both H53$\alpha$, and H66$\alpha$
lines are detected from only W51e2-W, the UC HII region. For the
H26$\alpha$, H$53\alpha$, and H66$\alpha$ lines from W51e2-W, we
integrated the line emission region with a size of $0.5\arcsec$ around
the H26$\alpha$ peak position.  The profiles of the H26$\alpha$,
H$53\alpha$ and H66$\alpha$ lines are multiplied by the frequency
ratios of $\nu_{H26\alpha}/\nu_{H26\alpha}$,
$\nu_{H26\alpha}/\nu_{H53\alpha}$ and
$\nu_{H26\alpha}/\nu_{H66\alpha}$, respectively, as shown in Figure
\ref{fig:profiles}. The measurements of the peak intensity
($I_{peak}$), radial velocity ($V_{LSR}$), and FWHM line width
($\Delta V$) for the H26$\alpha$, H$53\alpha$, and H66$\alpha$ lines
are given in Table \ref{tab:RRLs}.

\begin{figure*}[t]
\centering
\subfigure[]{\includegraphics[width=55mm,angle=-90]{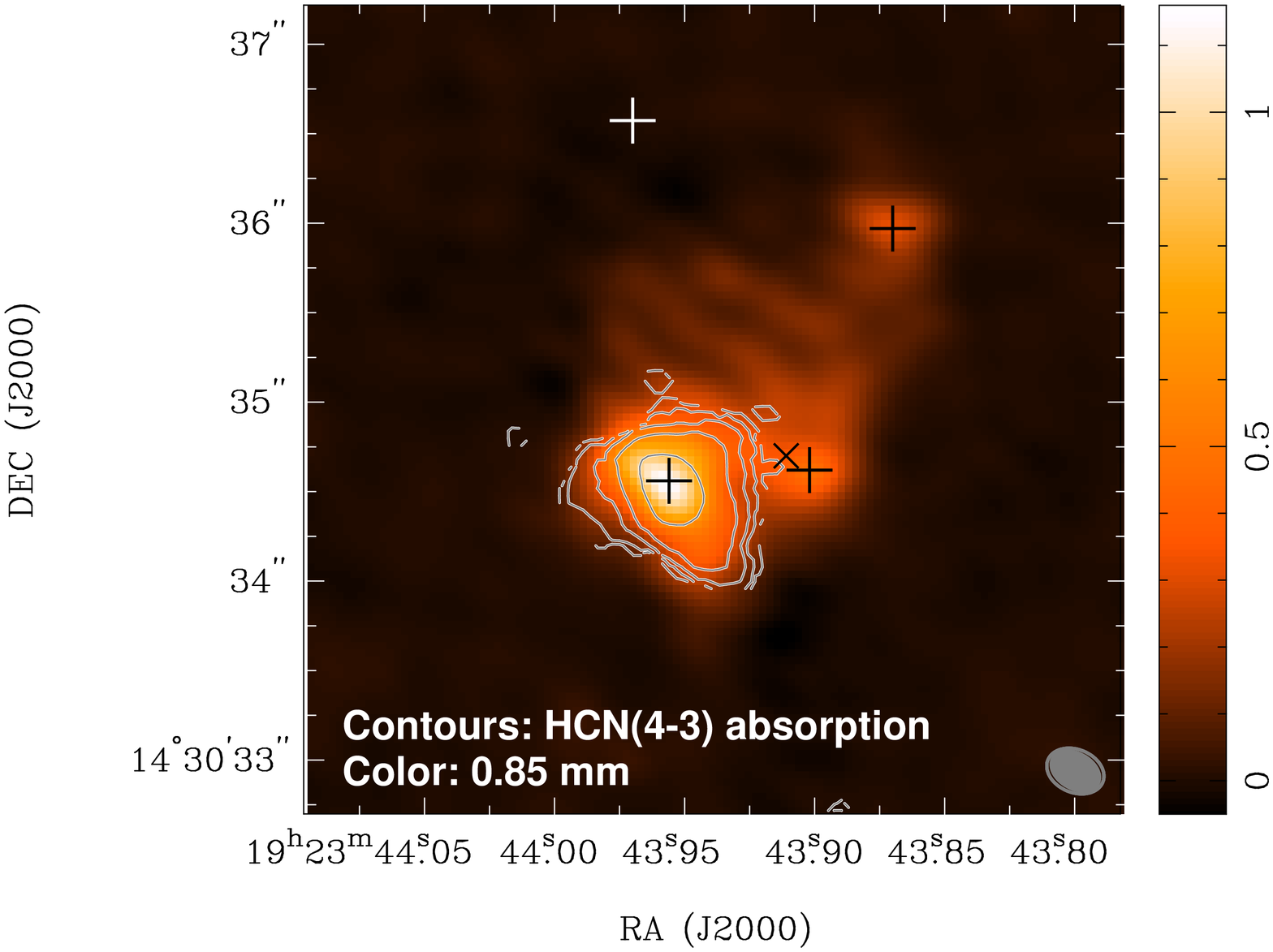}}
\subfigure[]{\includegraphics[width=55mm,angle=-90]{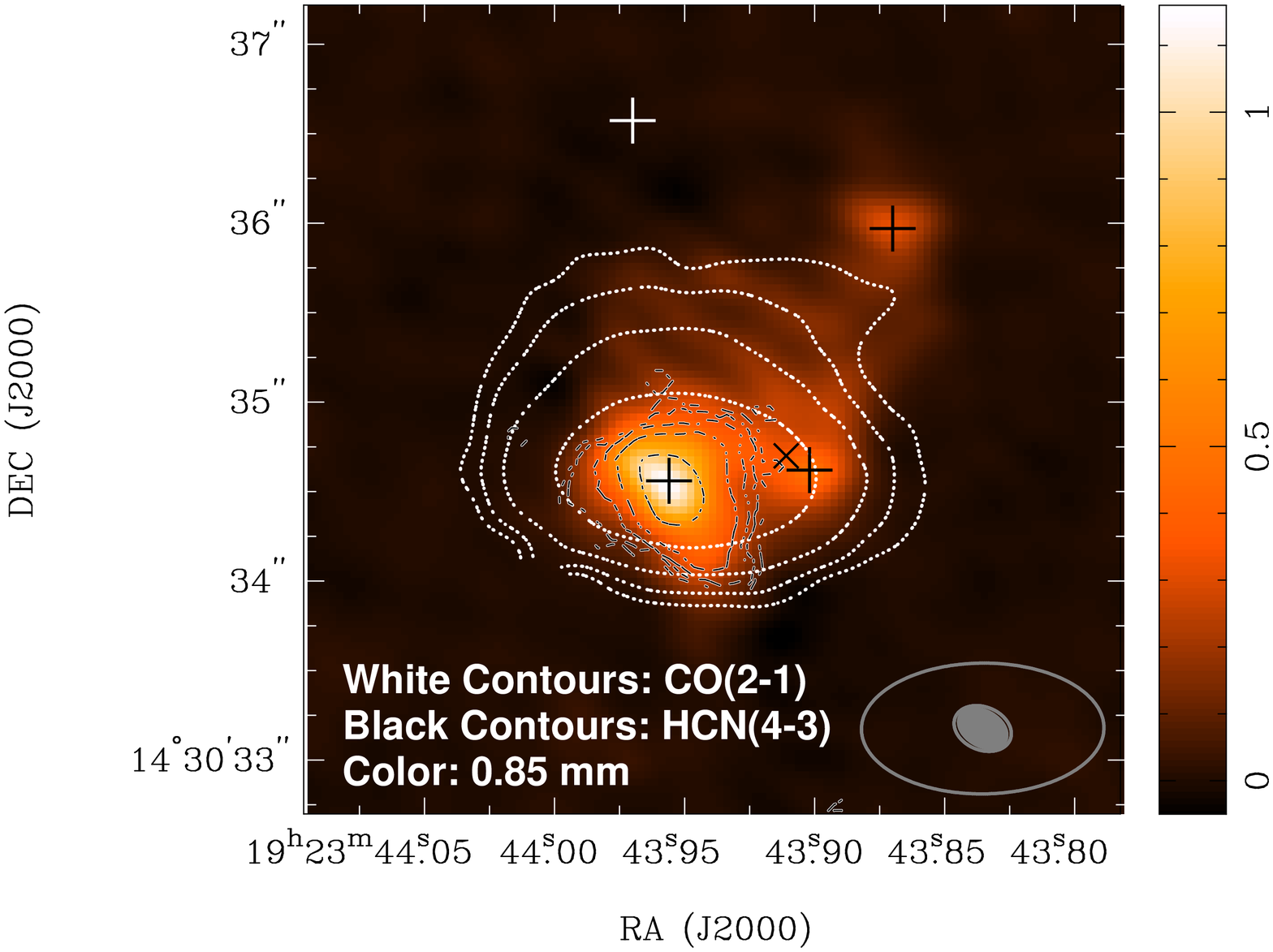}}\\
\subfigure[]{\includegraphics[width=40mm,angle=-90]{f4c.ps}}
\subfigure[]{\includegraphics[width=40mm,angle=-90]{f4d.ps}}
\subfigure[]{\includegraphics[width=40mm,angle=-90]{f4e.ps}}
\caption{
{\it(a)}: Image of the HCN(4-3) absorption line integrated from 44 to
62~km~s$^{-1}$. Dashed contours are $\pm 5\sigma \times 2^n$ ($n=0$,
1, 2, 3, ... and $\sigma=0.113$~Jy~beam$^{-1}$~km~s$^{-1}$), overlaid
on the gray-scaled continuum image at 0.85\,mm. The FWHM beam
$0.33\arcsec\times 0.24\arcsec$ (sup=0) is shown at bottom right.
{\it(b)}: Image of the CO(2-1) absorption line integrated from 43 to
71~km~s$^{-1}$. White dashed contours are $\pm 5\sigma \times 2^n$
($n=0$, 1, 2, 3, ... and $\sigma=0.771$~Jy~beam$^{-1}$~km~s$^{-1}$),
overlaid on the gray-scaled continuum image at 0.85\,mm.  The FWHM
beam $1.35\arcsec\times 0.73\arcsec$ (sup=0) is shown at bottom right.
{\it(c and d)}: The spectral profiles of the HCN(4-3) absorption line
toward W51e2-E and W51e2-W, respectively. The vertical dashed line
denotes the systematic velocity of 53.9$\pm$1.1\,km~s$^{-1}$.
{\it(e)}: The spectral profile of the CO(2-1) absorption line toward
W51e2-E, with a vertical dashed line for the systematic velocity.
}
\label{fig:molecular}
\end{figure*}

\begin{table*}
\small
\setlength{\tabcolsep}{0.8mm}
%
\caption{Core fluxes of W51e2 at Centimeter to Submillimeter bands}
\label{tab:SED}
\begin{tabular}{lccccl}
\hline
\hline
\multicolumn{1}{c}{parameters} &  {W51e2-W} & {W51e2-E} & {W51e2-NW} & {W51e2-N} & {References and Beamsize $\Theta$} \\
\hline
R.A.(J2000)& 19:23:43.90& 19:23:43.96& 19:23:43.87& 19:23:43.97& This paper\\
Decl.(J2000)& +14:30:34.62& +14:30:34.56& +14:30:35.97& +14:30:36.57& This paper\\
Deconvolved size ($\arcsec$)$^{\ast}$&  0.45   & 0.71        & 0.38        & 1.24        & This paper   \\
S$_{\rm 0.85mm}$(Jy)& 0.35$\pm$0.07& 3.30$\pm$0.20& 0.81$\pm$0.34&  &   This paper: $\Theta=0.4\arcsec$ \\
S$_{\rm 1.3mm}$(Jy)& & 2.15$\pm$0.12& 0.62$\pm$0.12& 0.73$\pm$0.08&  This paper: $\Theta=1.1\arcsec$ \\
S$_{\rm 7mm}$(Jy)& 0.62$\pm$0.09& 0.04$\pm$0.01& $<$ 0.0052& $<$ 0.0052&  This paper: $\Theta=0.4\arcsec$   \\
S$_{\rm 1.3cm}$(Jy)& 0.43$\pm$0.03& $<$ 0.004& $<$ 0.004&  $<$ 0.004&  This paper: $\Theta=0.4\arcsec$; \citet{GJW93}: upper limit \\
S$_{\rm 2cm}$(Jy)& 0.21$\pm$0.05& & & &  \citet{RWP90}: $\Theta=0.4\arcsec$   \\
S$_{\rm 3.6cm}$(Jy)& 0.067$\pm$0.004& $<$ 0.001& $<$ 0.001& $<$ 0.001&  \citet{GJW93}: $\Theta=0.21\arcsec$; \citet{GJW93}: upper limit    \\
S$_{\rm 6cm}$(Jy)& 0.025$\pm$0.003& --& --&  -- &  \citet{RWP90}: $\Theta=0.6\arcsec$   \\
\hline
\end{tabular}\\
\begin{tabular}{p{0.9\textwidth}}
$^{\ast}$The sizes of e2-W, e2-E, and e2-NW are derived from the
  0.85\,mm image, and e2-N is from the 1.3\,mm image.
\end{tabular}
\end{table*}

\subsection{ Absorption of HCN(4-3) and CO(2-1)}
The systematic velocity of W51e2 varies when it is measured using
different hot molecular lines \citep{SZH04,RSS04,ZHO98,RWP90}. In
order to minimize the possible effects of the different distributions
with different molecular lines, we averaged these measurements of
molecular lines H$^{13}$CO$^{+}$, SO$_2$, SiO, CH$_3$CN, and HCO$^{+}$
and obtained a mean value of 53.9$\pm$1.1\,km~s$^{-1}$. We adopt this
value as the systematic velocity for W51e2-E hereafter.

The lines of HCN(4-3) at $\nu_0=$ 354.505\,GHz and CO(2-1) at $\nu_0=$
230.538\,GHz were covered in the SMA observations at 0.85\,mm and
1.3\,mm, respectively. The dirty images were made with natural
weighting and deconvolved with the Hogbom Clark Steer hybrid
algorithm. The final rms noises of $\sigma=$ 65 and 94\,mJy
beam$^{-1}$ per channel (1\,km s$^{-1}$ in the channel width) were
achieved for HCN(4-3) and CO(2-1), respectively.  Figure
\ref{fig:molecular} shows the spectra of HCN(4-3) toward W51e2-E and
W51e2-W and the spectrum of CO(2-1) near the peak position of
W51e2-E. A significant absorption of the HCN(4-3) line is detected
against the submillimeter core, W51e2-E, and most of which is
redshifted with respect to the systematic velocity of 53.9$\pm$1.1\,km
s$^{-1}$ as indicated by the vertical dashed line in Figure
\ref{fig:molecular}(c). Note that the redshifted absorption of HCN
observed with the high angular resolution against the compact
continuum core corresponds to the cold gas in front of the dust core
moving toward it, i.e., the infall. The CO(2-1) spectrum shows a
broader redshifted feature in absorption than that of HCN. Due to the
poor angular resolution in the CO(2-1) observations, the observed
broad profile in absorption is caused not only by the infall but also
by the outflow. In particular, the high redshifted absorption arises
very likely from the outflow gas. In comparison with that of W51e2-E,
the spectrum toward W51e2-W only shows a weak spectral feature
(Figure \ref{fig:molecular}(d)) which can be characterized as an
inverse P Cygni profile, suggesting that only little molecular gas (if
present) falls onto the UC HII region.

To show the spatial distribution of the absorption gas, we carried out
a moment analysis. The calculation of the zeroth moment corresponds to
the integrated intensity over the velocity. In order to avoid
cancellation between emission and absorption spectral features in the
same beam area, we separate the absorption and emission in the moment
calculation. We found that both the blueshifted and redshifted
emission of the HCN(4-3) line (the images are not shown in this paper)
is extended from northwest to southeast across W51e2-E, which is, in
general, consistent with the molecular outflow direction as suggested
based on the CO(2-1) line observation \citep{KK08}.  The HCN(4-3)
absorption, integrated from 44 to 62\,km~s$^{-1}$ with a 5$\sigma$
cutoff in each channel (Figure \ref{fig:molecular}(a)), is found to be
mainly concentrated on W51e2-E. From the lower-resolution observation,
the distribution of the CO(2-1) absorption line integrated from 43 to
71\,km~s$^{-1}$ (a 5$\sigma$ cutoff in each channel, see Figure
\ref{fig:molecular}(b)), also shows the absorption peaks at W51e2-E
instead of W51e2-W. We also note that the CO(2-1) absorption shows a
relatively extended feature north of W51e2-E, further indicating that
the absorption of CO(2-1) might indeed be significantly contaminated
by the outflow gas.

Along with the absorption spectra, the distribution of the absorbing
gas in W51e2 evidently demonstrates that the submillimeter (dust) core
W51e2-E, instead of W51e2-W (the UC HII region), is the center of
accretion for the majority of the high-density gas from this molecular
core.

\subsection{Spectral Energy Distributions}
In lack of high angular resolution observations at
millimeter/submillimeter wavelengths, the flux densities from the
entire region of W51e2 were considered in the previous analyses of the
SED \citep{RWP90,SZH04,KZK08}. With the high-resolution observations
at the wavelengths from radio to submillimeter, we are now able to
spatially separate individual emission components in our SED analysis.

\begin{figure*}[t]
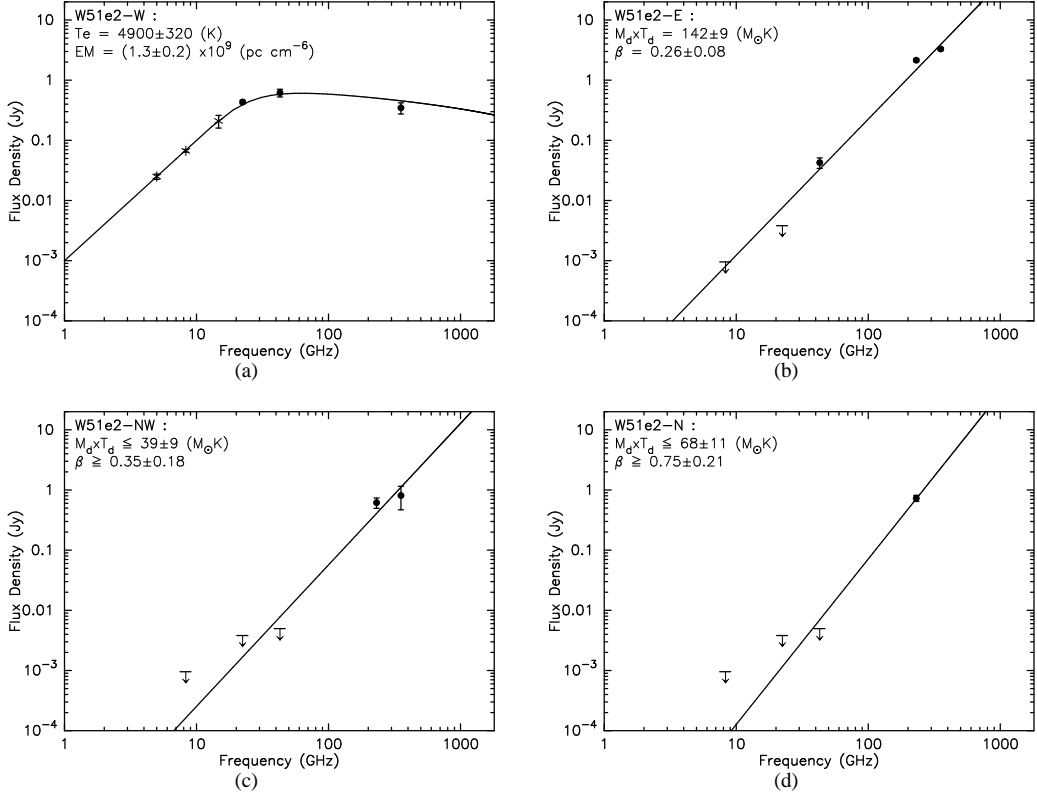

\centering
\subfigure[]{\includegraphics[angle=-90.0, width=65mm]{f5a.ps}}
\hspace{5mm}
\subfigure[]{\includegraphics[angle=-90.0, width=65mm]{f5b.ps}}\\
\subfigure[]{\includegraphics[angle=-90.0, width=65mm]{f5c.ps}}
\hspace{5mm}
\subfigure[]{\includegraphics[angle=-90.0, width=65mm]{f5d.ps}}
\caption{
The SED for W51e2 cores.
{\it (a)}: SED fitting for W51e2-W with a free$-$free emission
model. Black points denote the measurements from this work, and the
asterisks are from \citet{RWP90} and \citet{GJW93}. All flux values
are listed in Table \ref{tab:SED}.
{\it (b)}: The SED fitting for the W51e2-E core with a thermal dust
emission model. The arrows denote the upper limits from \citet{RWP90}
and \citet{GJW93}.
{\it (c)}: The SED fitting for the W51e2-NW core. The upper limit of
the flux density at 43\,GHz (7\,mm) is from our data (3$\sigma$ of the
7\,mm image).
{\it (d)}: The SED fitting for the W51e2-N core.
}
\label{fig:SED}
\end{figure*}

Adding the data at 3.6\,cm \citep{GJW93} and at 2 and 6\,cm
\citep{RWP90} along with the continuum data at the four wavelengths
(13, 7, 1.3, and 0.8\,mm) discussed in this paper, we have seven
measurements in flux densities covering nearly 2 order of magnitude in
frequency. For 13, 7, and 0.8 mm data, the flux density measurements
were made by fitting a Gaussian to the individual emission components
in the images with the same size of the convolved FWHM beam
(0.4$\arcsec$). For the 1.3\,mm data, we determined the flux density
by fitting multiple Gaussian components to the emission features in a
relatively lower resolution (1.1$\arcsec$), which results in
relatively large uncertainties. For the emission cores W51e2-W,
W51e2-E, W51e2-NW, and W51e2-N, the flux densities determined are
given in Table \ref{tab:SED}. The relevant continuum spectra for the
four emission components are shown in Figure \ref{fig:SED}.

\section{Nature of the W51e2 complex}
Four distinct components have been identified in the 3$\arcsec$ region
of W51e2 from the high-resolution images at wavelengths from
centimeter to submillimeter (Figure \ref{fig:350GHz} and
\ref{fig:con_rrls}). The nature of these components is discussed as
follows.

\subsection{W51e2-W: UC HII region}
W51e2-W is the only core detected at centimeter wavelengths, and its
continuum emission is dominated by the free$-$free emission from the
ionized core and traces the thermal HII region around the central
star. Figure \ref{fig:con_rrls}(b)$-$(f) show the UC HII region,
W51e2-W, including an unresolved core (a possible ionized disk) and a
northeast and southwest (NE$-$SW) extension as a possible outflow. The
NE and SW extensions are best to be seen at 7 and 13\,mm in Figure
\ref{fig:con_rrls}(b) and (c), which is in good agreement with the
previous observations at 3.6 and 1.3\,cm by
\citet{GJW93}. \citet{GJW93} found the spectral indices $\alpha$ of
about 2 (optically thick) and 0.4 for the NE and SW extensions of
W51e2-W, respectively, and suggested that there is a one-side (SW)
collimated ionized outflow from the core.

The H26$\alpha$ line is an excellent tracer for hyper-compact ionized
cores (star in Figure \ref{fig:con_rrls}(d)), providing a good
diagnosis for the presence of an ionized disk. On the basis of the SMA
observations with an angular resolution of 0.25$\arcsec$, the
hyper-compact ionized core which peaked at 59.1\,km~s$^{-1}$ in
W51e2-W has not been resolved, giving a limit on the intrinsic size of
$\theta_s<0.06\arcsec$ and linear size $<310$\,AU from a Gaussian
fitting. The elongation shown in the H53$\alpha$ image (Figure
\ref{fig:con_rrls}(e)) agrees with the radio continuum images observed
at 1.3\,cm \citep[Figure \ref{fig:con_rrls}(c) and][]{GJW93}, 3.6\,cm
\citep{GJW93}, and 7\,mm (Figure \ref{fig:con_rrls}(b)), suggesting
that the NE$-$SW extension observed in the H53$\alpha$ line
corresponds to the expansion of ionized gas (or outflow). The
H66$\alpha$ line emission also has a similar elongation (Figure
\ref{fig:con_rrls}(f)), with peak intensity near Gaume's position in
the NE and an extension structure in the SW. We also found a
significant velocity gradient of the H53$\alpha$ line in the W51e2-W
region, in agreement with what was observed by \citet{KK08},
redshifted in the SW and blueshifted in the NE. The broad wings of the
recombination lines shown in Figure \ref{fig:profiles} (also in Table
\ref{tab:RRLs}) corresponding to extended, optically thin emission
indicate that it is hard to explain them as an ionized disk. The
velocity gradient observed in the H53$\alpha$ line is more likely
produced by an ionized outflow rather than an ionized accretion disk.

We fitted the SED of W51e2-W with a free$-$free emission model
     \begin{eqnarray}
       S_{\nu} &=& \Omega_s B_{\nu}(T_e)(1-e^{-\tau_c})
               ~~~~~({\rm Jy}),    \label{eq:Flux_HII}
     \end{eqnarray}
where $S_{\nu}$ is the flux density at frequency $\nu$, $\Omega_s$ is
the solid angle of the source, $B_{\nu}(T_e)$ is the Planck function,
and $T_e$ is the electron temperature. The continuum optical depth is
expressed by \citep{MH67} as
     \begin{eqnarray}
       \tau_c &=& 0.0824\left(\frac{T_e}{{\rm K}}\right)^{-1.35}
                  \left(\frac{\nu}{{\rm GHz}}\right)^{-2.1} 
                  \left(\frac{\rm EM}{{\rm pc~cm^{-6}}}\right)\alpha(\nu, T_e).
                                   \label{eq:tau_c}
     \end{eqnarray}
The emission measure (EM) can be written as
     \begin{eqnarray}
       {\rm EM} = N_e^2L f_V,              \label{eq:EM}
     \end{eqnarray}
where $N_e$ is the mean electron density, $L$ is the path length, and
$f_V$ is the volume filling factor. The dimensionless factor
$\alpha(\nu, T_e)$ is the order of unity \citep{MH67}. For a
homogeneous ionized core, the model can be expressed with two free
parameters, namely, $T_e$ and EM, for the first approximation
described above.

The solid curve in Figure \ref{fig:SED}(a) shows the best fit to the
data, suggesting a turnover frequency of $\nu_{0}\approx27$\,GHz.  The
derived mean values for the physical parameters for the overall HII
region, $T_e$=4900$\pm$320\,K and
EM=(1.3$\pm$0.2)$\times10^{9}$\,pc~cm$^{-6}$, are similar to the
previous result in \citet{ZHO98}. Thus, the mean electron density of
$N_e\sim$3$\times$10$^{5}$\,cm$^{-3}$ is inferred on the assumption of
$Lf_V\sim$0.5$\arcsec\times$5.1\,kpc.

Both the Lyman continuum flux ($N_L$) and the excitation parameter
($U$) can be derived \citep{Rub68,Pan73,MES76} as follows:
     \begin{eqnarray}
       N_{L} &\ga& 7.5 \times 10^{46} {\rm s^{-1}}\left(\frac{S_{\nu}}{\rm Jy}\right)
                 \left(\frac{D}{\rm kpc}\right)^{2}
                  \left(\frac{\nu}{\rm GHz}\right)^{0.1} \times \nonumber\\
             &&  \left(\frac{T_e}{10^4{\rm K}}\right)^{-0.45},
                   \label{eq:NL} \\
       U &=& 3.155\times10^{-15} {\rm pc~cm^{-2}}
                                   \left(\frac{N_L}{\rm s^{-1}}\right)^{\frac{1}{3}}
                                   \left(\frac{T_e}{10^4{\rm K}}\right)^{\frac{4}{15}},
     \end{eqnarray}
where $D$ is the distance to the source. We assumed $\alpha(\nu,
T_e)\sim1$. Based on our model fitting, the continuum emission from
e2-W becomes optically thin at a frequency $>$80\,GHz ($\tau<$0.1),
and we obtained $N_L$=3.0$\times$10$^{48}$ s$^{-1}$ and
$U$=37.6\,pc~cm$^{-2}$. The inferred Lyman continuum flux requires a
massive star equivalent to a zero-age mean-sequence star of type O8
located inside the UC HII region W51e2-W \citep{Pan73}, which is
consistent with the result of O7.5 in \citet{RWP90}. Alternatively, a
cluster of B-type stars can also be responsible for the
ionization. The total ionized mass within 0.5$\arcsec$ of W51e2-W is
$\sim$0.02\,M$_{\sun}$, which agrees with the value derived by
\citet{GJW93}.

For the hyper-compact core ($\theta_s<0.06\arcsec$) as observed with
the H26$\alpha$ line, the electron temperature can be estimated from
the H26$\alpha$ line and the continuum flux density at 354\,GHz
(0.85\,mm) on the assumption of optically thin, LTE condition:
\begin{eqnarray}
T_e^*&=&\Bigg[\left(\frac{6985}{\alpha(\nu,T_e)}\right)
     \left(\frac{\nu}{\rm GHz}\right)^{1.1}
     \left(\frac{\Delta V_{\rm H26\alpha}}{\rm km~s^{-1}}\right)^{-1}
     \left(\frac{S_{\rm 0.85\,mm}}{S_{\rm H26\alpha}}\right) \times \nonumber \\
     &&\left(\frac{1}{1+\frac{N(He)}{N(H)}}\right)\Bigg]^{0.87}.
\end{eqnarray}
Assuming $N(He)/N(H)=0.096$ \citep{Meh94} and $\alpha(\nu,T_e)=1$, we
have $T_e^*=12,000\pm2,000$\,K which appears to be considerably higher
than the mean electron temperature of the overall HII region. The EM
of the hyper-compact HII core can be assessed as
\begin{eqnarray}
{\rm EM} &=& 7.1~{\rm pc~cm^{-6}} \left(\frac{T_e}{\rm K}\right)^{3/2}
                    \left(\frac{\theta_s}{\rm arcsec}\right)^{-2}
                    \left(\frac{\lambda}{\rm mm}\right)
                    \left(\frac{S_{L}}{\rm Jy}\right)\times \nonumber \\
                    &&\left(\frac{\Delta V}{\rm km~s^{-1}}\right), \label{eq:rrlEM}
\end{eqnarray}
where $T_e$ is the electron temperature, $\theta_s$ is the intrinsic
size of the HII region, $S_L$ is the peak line intensity in Jy,
$\Delta V$ is the FWHM line width in km~s$^{-1}$, and $\lambda$ is the
observing wavelength in millimeters.  Based on the measurements of the
H26$\alpha$ line from the hyper-compact ionized core together with the
assumption of $T_e\approx T_e^*$, EM $> 7\times10^{10}$ pc cm$^{-6}$
is found.  The corresponding lower limit of the volume electron
density is $N_e> 7\times10^{6}$\,cm$^{-3}$ assuming
$Lf_V<0.06\arcsec\times5.1$ kpc. The inferred high electron
temperature and high electron density suggest that the H26$\alpha$
line arises from a hot, very compact region ($< 0.06\arcsec$) which is
probably close to the central ionizing star(s). With the nature of
optically thin and sensitive to the high-density ionized gas, the
H26$\alpha$ line is an excellent tracer of the hot, high-density
ionized region surrounding the ionizing source. In comparison with the
mean electron temperature ($T_e$=4900\,K) inferred from the
free$-$free emission in a large area (0.5$\arcsec$), the high
temperature of $T_e$=12,000\,K derived from the H26$\alpha$ in a
compact area ($< 0.06\arcsec$) suggests that a temperature gradient is
present along the radius of the UC HII region W51e2-W.

Observations of RRLs in a wide range at wavelengths from centimeters
to submillimeters also offer a means to further explore physical
conditions of W51e2-W. Table \ref{tab:RRLs} summarizes the reliable
measurements of three RRLs from W51e2, namely, H26$\alpha$,
H53$\alpha$, and H66$\alpha$.  We plotted the spectral profiles of the
three RRLs together, and each of the profiles has been multiplied by a
factor of $\nu_{26\alpha}/\nu_{n\alpha}$, the ratio of the H26$\alpha$
line frequency to that of RRLs (Figure \ref{fig:profiles}). In the
case of optically thin and no-pressure broadening, the three line
profiles should match each other. Comparing H53$\alpha$ with
H26$\alpha$, we find that the peak intensities of the two lines agree
with each other, which suggests that, in general, both the H53$\alpha$
and H26$\alpha$ lines are under an optically thin, LTE
condition. Several velocity peaks in the H26$\alpha$ line indicate
multiple kinematic components in the hyper-compact HII core.  The
H53$\alpha$ line is characterized by a relatively smooth profile with
large velocity wings. The large velocity wings observed in the
H53$\alpha$ line can be explained by the large velocity gradient in
the lower-density electron gas of the ionized outflow \citep{GJW93}.
However, the line intensity of H66$\alpha$ appears to be significantly
weaker than the other two high-frequency RRLs. The peak of the
modified line profile of H66$\alpha$ is a factor of $\sim 5$ less than
that of the other two. Considering the fact that the frequency of
H66$\alpha$ is below the turnover frequency of
$\nu_{0}\approx27$\,GHz, we found that the H66$\alpha$ line from the
ionized gas is obscured severely due to the self-absorption process in
the hyper-compact HII core. From Equation (2), a mean optical depth of
$\tau_c=1.6$ at $\nu_{H66\alpha}=22.36$ GHz is found. The exponential
attenuation of $exp(-1.6)\approx0.2$ suggests that the H66$\alpha$
line is attenuated mainly due to the self-absorption in the ionized
core. In addition, the pressure broadening effect which weakens the
lower frequency lines needs to be assessed.

\begin{table}[t]
\small
\centering
\tablewidth{0pt}
\setlength{\tabcolsep}{0.8mm}
%
\caption{Parameters of RRLs derived from our data$^{\ast}$}
\label{tab:RRLs}
\begin{tabular}{lccc}
\hline
\hline

\multicolumn{1}{l}{} & {${\rm H26\alpha}$} &
{${\rm H53\alpha}$} & {${\rm H66\alpha}$} \\
Rest Frequency (GHz) & 353.623& 42.952& 22.364   \\
\hline
Measurements \\
$I_{peak}$ (Jy) & 1.29$\pm$0.02& 0.137$\pm$0.001& 0.015$\pm$0.003 \\
$V_{LSR}$ (km~s$^{-1}$)& 59.1$\pm$0.2& 60.4$\pm$0.2& 62.1$\pm$1.3 \\
$\Delta V$ (km~s$^{-1}$)& 23.1$\pm$0.5& 32.0$\pm$0.3& 35.1$\pm$ 8.3 \\
\hline
Derived line broadenings \\
$\Delta V_P$ (km~s$^{-1}$)& 0.01& 1.3& 6.5  \\
$\Delta V_T$ (km~s$^{-1}$)& 15.0  & 15.0 & 15.0  \\
$\Delta V_t$ (km~s$^{-1}$)& 17.5& 28.3& 31.1 \\
$\Delta V_D$ (km~s$^{-1}$)& 23.1& 32.0& 34.5 \\
\hline
\end{tabular}\\
\begin{tabular}{p{0.47\textwidth}}
$^{\ast}$All line profiles are integrated from the
  0.5$\arcsec\times$0.5$\arcsec$ region around the peak position of
  H26$\alpha$ from maps of the same beamsize of 0.25$\arcsec$.
\end{tabular}
\end{table}

The total line width of RRLs ($\Delta V$) can be expressed by the
Doppler broadening ($\Delta V_D$) and pressure broadening ($\Delta
V_P$). According to \citet{BL71} and \citet{GS02}, we have
     \begin{eqnarray}
       \Delta V_D&=&{\rm km~s^{-1}}
                  \sqrt{0.0458\times\left(\frac{T_e}{\rm K}\right)
                    +\left(\frac{\Delta V_t}{\rm km~s^{-1}}\right)^{2}},\\
       \Delta V_P&=&
                  3.74\times10^{-14}~{\rm km~s^{-1}}~n^{4.4}
                  \left(\frac{\lambda}{\rm mm}\right)
                   \left(\frac{N_e}{\rm cm^{-3}}\right) \times \nonumber \\
                 &&\left(\frac{T_e}{\rm K}\right)^{-0.1}, \label{eq:V_P} \\
       \Delta V&=& \sqrt{\Delta V_{D}^2+\Delta V_P^2},
     \end{eqnarray}
where $\Delta V_t$ is the turbulent broadening of the gas including
broadening due to outflows and disk rotations and $n$ is the principal
quantum number of the transition. We calculated broadenings on the
mean electron temperature ($T_e\approx$4900\,K) and the mean electron
density ($N_e\approx$3$\times$10$^5$\,cm$^{-3}$) of the UC HII region.
The thermal broadening ($\Delta V_T$) is 15\,km~s$^{-1}$ for
$T_e$=4900\,K. The pressure broadening is a strong function of the
principal quantum number, i.e., $\Delta V_P$ increases drastically
toward the low-frequency lines. For the high-frequency line
(H26$\alpha$), the pressure broadening ($\Delta
V_P=0.01$\,km~s$^{-1}$) can be ignored, while for the H66$\alpha$
line, the pressure broadening of $\Delta V_P$=7\,km~s$^{-1}$ becomes
significant. The pressure broadening is $\Delta V_P$=76\,km~s$^{-1}$
for the H92$\alpha$ line, which can actually wash out the profile of
the line emission. Thus, in addition to the opacity effect, the
pressure broadening may also make a considerable contribution to
diminish the H92$\alpha$ line which was not detected in the
observations of \citet{Meh94}. The turbulent and/or the dynamical
motions of the ionized gas contribute a total of $\Delta
V_t\approx$18\,km~s$^{-1}$ in the line broadening for the H26$\alpha$
line and $\Delta V_t\approx$28 and 31\,km~s$^{-1}$ for the H53$\alpha$
and H66$\alpha$ lines, respectively.
The H26$\alpha$ line traces the high-density ionized gas in the
hyper-compact core or the ionized disk (if it exists) and is less
sensitive to the lower-density outflow which produces the broad wings
in the line profiles.  The turbulent and/or dynamical broadening
becomes dominant in the H53$\alpha$ and H66$\alpha$ lines, and these
lower frequency RRLs (H53$\alpha$ and H66$\alpha$) better trace the
lower-density electron gas in the ionized outflow.

In short, W51e2-W consists of a hyper-compact HII core ($<310$\,AU)
with an emission measure of EM $> 7\times10^{10}$\,pc~cm$^{-6}$ and an
ionized outflow.  An early-type star equivalent to an O8 star (or a
cluster of B-type stars) is postulated to have formed within the
hyper-compact HII core. The broadening of the line profiles is
dominated by the Doppler broadening including both thermal and
turbulence/dynamical motions. Pressure broadening becomes significant
only for the H66$\alpha$ line and the lines with a larger principal
quantum number.

\subsection{W51e2-E: A massive proto-stellar core}

W51e2-E is located $\sim0.9\arcsec$ east of the UC HII region
W51e2-W. It is the brightest source in the 0.85\,mm continuum image
(Figure \ref{fig:350GHz}). The flux density drops drastically at 7\,mm
(Figure \ref{fig:con_rrls}(b)). It cannot be detected at longer
wavelengths. The SED of the continuum emission from W51e2-E (Figure
\ref{fig:SED}(b)) appears to arise from a dust core associated with
proto-stars.

With the Rayleigh$-$Jeans approximation, the continuum flux density
from a homogeneous, isothermal dust core can be described as
\citep{LH97}
\begin{eqnarray}
  S_d(\nu) &=& 6.41 \times10^{-6} {\rm Jy}
                              \left(\frac{\kappa_0}{\rm cm^2g^{-1}}\right)
                              \left(\frac{\nu}{\nu_0}\right)^{\beta}
                              \left(\frac{\nu}{\rm GHz}\right)^2 \times \nonumber \\
                             && \left(\frac{M_d}{\rm M_{\sun}}\right)
                                \left(\frac{T_d}{\rm K}\right)
                                \left(\frac{D}{\rm kpc}\right)^{-2},
         \label{eq:Flux_dust}
\end{eqnarray}
where $\kappa_0$ is the dust opacity at frequency $\nu_0$, $\beta$ is
the power-law index of the dust emissivity, $M_d$ is the mass of dust,
$T_{d}$ is the dust temperature in K, and $D$ is the distance to the
source. We adopt the dust opacity value of
$\kappa(\lambda=1.3mm)=0.8~{\rm cm^2g^{-1}}$ \citep{OH94,LH97} and
have three free parameters for the dust core ($M_{d}$, $T_{d}$, and
$\beta$).

The solid line in Figure \ref{fig:SED}(b) shows the best least-squares
fitting to the observed data. The index of $\beta=0.26\pm0.08$ derived
from our fitting indicates that emission from the e2-E core might be
dominated by large grains of dust \citep{MN93,CZO95,AW07}. In
addition, because of a lack of flux density measurements at higher
frequencies or shorter wavelengths, the dust mass and temperature are
degenerate in our best fitting, i.e., $M_dT_d=142\pm9$\,M$_{\sun}$K.
\citet{RSS04} derived the kinetic temperature in W51e2 to be
$153\pm21$\,K from the high-resolution observations of CH$_3$CN, which
is close to the dust temperature $T_d$=100\,K used in
\citet{ZHO98}. Adopting that $T_d=$100\,K and assuming that the ratio
of H$_2$ gas to dust is 100, we found that the total mass of
$\sim$140\,M$_{\sun}$ is in the W51e2-E core. Emission from the
W51e2-E region has been slightly resolved, so that the dust core of
W51e2-E might host a number of proto-stellar cores which accrete the
surrounding molecular gas as indicated by both the HCN and CO
absorption lines (see Section 2.4). This proposed scenario for W51e2-E
is also consistent with the hourglass-like magnetic fields inferred
from polarization measurements of \citet{THK09} for the W51e2
region. The configuration center of $B$ vectors coincides with the
peak position of the W51e2-E dust core.

From a Gaussian fitting to the absorption spectrum of the HCN(4-3)
toward the center of W51e2-E (Figure \ref{fig:molecular}(c)), we
determined the peak line intensity $\Delta
I_L=-1.3\pm0.1$\,Jy~beam$^{-1}$, the line center velocity $V_{HCN}$
=$56.4\pm0.2$\,km~s$^{-1}$ and the line width of $\Delta
V_{HCN}$=$9.5\pm0.4$\,km~s$^{-1}$. The infall velocity of the gas is
the offset between the center velocity and the systematic velocity,
i.e., $V_{in}=V_{HCN}-V_{sys}\approx2.5$\, km~s$^{-1}$.  The peak
continuum intensity of W51e2-E at 0.85\,mm is
$I_C=1.2\pm0.01$\,Jy~beam$^{-1}$ (corresponding to a brightness
temperature of $144\pm1$\,K). The ratio of $-\Delta I_{L}/I_C\sim1$
suggests that the absorption line is saturated.  Thus, a lower limit
on the optical depth \citep[see][]{QZM08} is $\tau_{HCN}>3$. Assuming
that the excitation temperature of HCN(4-3) equals the dust
temperature, 100\,K, we obtained a lower limit of HCN column density
of $7.9\times10^{15}$\,cm$^{-2}$. Taking
$[H_2]/[HCN]\sim0.5\times10^{8}$ \citep{IGH87} and the size of the
infall region as 0.75$\arcsec$ ($\sim4000$\,AU, see Figure
\ref{fig:molecular}(a)), we found that the hydrogen volume density
$N_{H_2}\sim6.9\times10^6$\,cm$^{-3}$.  The infall rate of the gas can
be estimated by
\begin{eqnarray}
  {\rm d}M/{\rm d}t
              &=& 2.1\times10^{-5} {\rm M_{\sun}yr^{-1}}
                \left(\frac{\theta_{in}}{\rm arcsec}\right)^2
                \left(\frac{D}{\rm kpc}\right)^2
                \left(\frac{V_{in}}{\rm km~s^{-1}}\right) \times \nonumber \\
              & &
                \left(\frac{N_{H_2}}{\rm 10^6 cm^{-3}}\right),
         \label{eq:dMdt}
\end{eqnarray}
where $2\theta_{in}=0.75\arcsec$ is the diameter of the infalling
region, $D$ is the distance to the source, $V_{in}$ is the infall
velocity, and the molecular mass ratio $m/m_{H_2}=1.36$ is
assumed. With the derived parameters a lower limit of infall rate
$1.3\times10^{-3}$ M$_{\sun}$~yr$^{-1}$ is inferred, suggesting that
W51e2-E is the accretion center of the W51e2 complex.

\subsection{W51e2-NW and W51e2-N}
W51e2-NW was clearly detected by \citet{THK09} in the continuum
emission at 0.87\,mm. They also found a significant concentration of
$B$ fields at the e2-NW. \citet{GDS81} and \citet{IWO02} detected
bright and compact H$_2$O masers near this source. We also detected
local enhanced continuum emission at 0.85 and 1.3\,mm. No significant
continuum emission has been detected at 7 and 13\,mm
(Figure\ref{fig:350GHz} and \ref{fig:con_rrls}), and there are no
significant absorption lines of HCN(4-3) and CO(2-1) in this region.
We fitted the SED of continuum emission with a thermal dust model as
shown in Figure \ref{fig:SED}(c). A lower limit of $\beta \geq 0.35$
is derived from the fitting, which leads to an upper limit of $M_dT_d
\leq 39$\,M$_{\sun}$K. Assuming the dust temperature in this region to
be $\sim$100\,K, we inferred an upper limit of 40\,M$_{\sun}$ in the
total gas mass if the ratio of H$_2$ gas to dust is equal to
100. W51e2-NW appears to be a massive core at a very early phase of
star formation.

Located $\sim2\arcsec$ north from W51e2-E, W51e2-N is detected in
continuum emission from our lower-resolution
(1.4$\arcsec\times0.7\arcsec$) image at 1.3\,mm (Figure
\ref{fig:con_rrls}(a)). This source appears to be resolved out with a
high angular resolution (0.3$\arcsec\times0.2\arcsec$) observation at
0.85\,mm. No significant continuum emission was detected at the longer
wavelengths. The SED consisting of the flux density at 1.3\,mm and the
upper limits at 7, 13, and 36\,mm is also fitted with a thermal dust
model. From the fitted values of the parameters $\beta \geq 0.75$ and
$M_dT_d \leq 68$\,M$_{\sun}$K, we found an upper limit of $M_d\la
70$\,M$_{\sun}$ if $T_d\sim$100\,K and the H$_2$ gas to dust ratio is
100. No maser activities have been detected from W51e2-N, suggesting
that W51e2-N is probably a primordial molecular clump in W51e2.

\subsection{A propagating scenario of star formation in W51e2}
Apparently, in the massive molecular core W51e2, the gravitational
collapse occurred first at W51e2-W where an O8 star or a cluster of
B-type stars were formed. Based on the absorption spectrum of the HCN
line, there appears to be little molecular gas present in the
immediate environs.  The accretion appears to be paused by the
intensive radiation pressure from the central ionized star(s). Most of
the surrounding gas ($\sim$0.02\,M$_{\odot}$) has been ionized. The
thermal pressure in the UC HII region drives an ionized outflow from
the ionized core. The submillimeter observations of dust emission and
molecular line absorption show that the major accretion now has been
re-directed to W51e2-E, 0.9$\arcsec$ ($\sim$4600\,AU) east from
W51e2-W. W51e2-E becomes the new dominant gravitational center
accreting mass from the surroundings with a rate of $10^{-3}$
M$_\odot$ yr$^{-1}$. Star formation activities take place in this
massive core, as shown by maser activities, outflow, and organized
$B$-field structure. Since no free$-$free emission and no RRLs have
been detected, the massive core W51e2-E likely hosts one or more
massive proto-stars. The offset of the radial velocities between the
two cores, W51e2-E and W51e2-W, is $\delta
V=59.1-53.9=5.2$\,km~s$^{-1}$. If W51e2-W and W51e2-E are
gravitationally bound and W51e2-W is circularly orbiting around
W51e2-E, W51e2-E has a dynamical mass of $>140$ M$_{\odot}$, in good
agreement with the mass derived from the SED analysis.

A bipolar outflow from W51e2-E has been suggested from the
observations of molecular lines. The impact of the outflow on the
medium in its path may trigger the collapse of the sub-core W51e2-NW
and induce further star formation activities there. W51e2-N represents
a gas clump with a considerable amount of mass for the star
formation. This speculative scenario appears to reasonably explain
what we have observed in the W51e2 complex.

\section{Summary and conclusions}
We have presented high-resolution images of the W51e2 region at 0.85,
1.3, 7, and 13\,mm for continuum and hydrogen recombination lines
(H26$\alpha$, H53$\alpha$, and H66$\alpha$) and the molecular lines
(HCN(4-3) and CO(2-1)). The W51e2 complex has been resolved into four
distinct components, W51e2-W, W51e2-E, W51e2-NW, and W51e2-N. We have
carried out a comprehensive analysis of the continuum SED, the RRLs,
and the molecular absorption lines and found that the four cores are
at different phases of massive star formation.

1. W51e2-W, associated with the UC HII region, is the only source from
which the H26$\alpha$, H53$\alpha$, and H66$\alpha$ lines have been
detected. The Lyman continuum flux inferred from the SED analysis
suggests that a massive O8 star or a cluster of B-type stars have been
formed within the HII region. The unresolved H26$\alpha$ line emission
region suggests the presence of a hot ($T_e^*=12,000\pm 2,000$ K),
hyper-compact ionized core with a velocity of
$59.1\pm0.2$\,km~s$^{-1}$, a linear size of $<310$\,AU, and a large
emission measure EM$>7\times10^{10}$\,pc~cm$^{-6}$ (or a high density
$N_e>7\times10^6$ cm$^{-3}$). The H53$\alpha$ and H66$\alpha$ line
images show an SW elongation from the hyper-compact ionized core,
suggesting either an expansion or an outflow of the ionized gas with
relatively lower density. No significant detection of the HCN
molecular line in absorption against the compact HII region indicates
that the significant accretion onto this core has been stopped.  The
line ratios between the H26$\alpha$, H53$\alpha$, and H66$\alpha$ show
that both the H26$\alpha$ and H53$\alpha$ lines from W51e2-W are in an
optically thin, LTE condition. The emission of the H66$\alpha$ line
from the hyper-compact core appears to be attenuated mainly due to the
self-absorption in this HII region. The line profile of the
H26$\alpha$ appears to be dominated by the Doppler broadening due to
the thermal motions of the hot electrons while the line broadening in
both the H53$\alpha$ and H66$\alpha$ is dominated by the Doppler
effect due to the dynamical motions of the ionized outflow. The
pressure broadening in both the H26$\alpha$ and H53$\alpha$ lines is
negligible and becomes significant for the H66$\alpha$ line
corresponding to $\Delta V_P\sim$ 7\,km s$^{-1}$.

2. W51e2-E is a massive ($\sim$140\,M$_{\sun}$) dust core as suggested
based on the SED analysis. No hydrogen recombination lines and no
radio continuum emission ($\lambda>1$\,cm) have been detected from
this region, which suggests that W51e2-E is the major core hosting
massive proto-stellar objects. Both the absorptions of HCN(4-3) and
CO(2-1) are redshifted with respect to the systematic velocity,
$53.9\pm1.1$\,km~s$^{-1}$, indicating that a large amount of molecular
gas moves toward the massive core. From the ratios of the absorption
line to continuum, a large infall rate of
$>1.3\times10^{-3}$\,M$_{\sun}$yr$^{-1}$ is inferred from the HCN
absorption line, corresponding to the accretion radius of 2000 AU.

3. W51e2-NW appears to be a massive ($\la$ 40\,M$_{\sun}$) core at an
earlier phase of star formation.
Given its location in the projected path of the molecular outflow from
W51e2-E, the star formation activities of W51e2-NW are likely to be
triggered by the outflow impact on the medium in the W51e2 complex.

4. W51e2-N appears to be a relatively less dense but massive clump
($\la$ 70\,M$_{\sun}$) from which no obvious evidence has been found
for star formation activities.

\acknowledgments
We thank the referee for helpful comments, Dr. S.-L. Qin for providing
help with the data reduction and analysis, and Professor Y.-F. Wu for
helpful discussions. H. Shi and J.-L.  Han are supported by the
National Natural Science Foundation (NNSF) of China (10773016,
10821061, and 10833003) and the National Key Basic Research Science
Foundation of China (2007CB815403).


\bibliographystyle{apj}
\bibliography{ms}






\end{document}